\documentclass[12pt]{article}
\usepackage{graphicx}
\usepackage{amsmath,amssymb,amsfonts}

\DeclareMathAlphabet\mathbfcal{OMS}{cmsy}{b}{n}

\begin{document}

\title{Interaction of light with gravitational waves}

\author{Shahen Hacyan
}

\renewcommand{\theequation}{\arabic{section}.\arabic{equation}}

\maketitle
\begin{center}

{\it  Instituto de F\'{\i}sica,}

{\it Universidad Nacional Aut\'onoma de M\'exico,}

{\it Apdo. Postal 20-364, M\'exico D. F. 01000, Mexico.}

%e-mail: hacyan@fisica.unam.mx

\end{center}
\vskip0.5cm

\begin{abstract}

The physical properties of  electromagnetic waves in the presence of a gravitational plane wave are analyzed.
Formulas for the Stokes parameters describing the polarization of light are obtained in closed form. The
particular case of a constant amplitude gravitational wave is worked out explicitly.

\end{abstract}


%Key words: .

 \maketitle

\newpage

\section{Introduction}

The propagation of electromagnetic waves in a gravitational field is an interesting problem in general, and it is
particularly relevant to the detection of gravitational waves by interferometric methods \cite{ligo} or
polarization of the cosmic microwave background \cite{rus,cosm}. Previous works on the subject started with
Plebanski's article on the scattering of electromagnetic waves by weak gravitational fields \cite{pleb}. Exact but
purely formal solutions of Einstein's equations for interacting electromagnetic and gravitational waves were
obtained by Sibgatullin \cite{sib}. More recently, electromagnetic waves in the background of a gravitational wave
(described by the Ehlers-Kundt metric \cite{ehler}) were analyzed by the present author \cite{hac}.

The aim of the present paper is to work out a general formalism which describes the polarization of light produced
by a gravitational wave. The formalism, as developed in Section 2, is based on the formal equivalence between an
anisotropic medium and a gravitational field; a general formula is thus obtained for the Stokes parameters. As an
example of application, a constant amplitude gravitational wave is considered in Section 3 and it is
shown that it produces a linear polarization of light.

\section{Electromagnetic and gravitational field}

The metric of a plane gravitational field propagating in the $z$ direction is (see, e.g., \cite{LL})
\begin{equation}
ds^2 =-dt^2 + (1+f)~ dx^2 +(1-f)~dy^2 +2 g ~dx~dy +dz^2,\label{1}
\end{equation}
where $f(u)$ and $g(u)$ are functions of the null coordinate $u=t-z$ (in this article we set $c=1$). Since the
gravitational field is assumed to be weak, only first  order terms in $f$ and $g$ need be considered.

The Maxwell equations in curved space-time are formally equivalent to these same equations in flat space-time,
with electric  and magnetic polarizations ${\bf P}$ and ${\bf M}$ due to the gravitational field \cite{pleb}. The
relations between electric and magnetic field vectors ${\bf E}$ and ${\bf H}$, and electric displacement and
magnetic induction ${\bf D}$ and ${\bf B}$ are the usual ones,
\begin{equation}
{\bf D} = {\bf E} + 4 \pi {\bf P} , \quad {\bf H} = {\bf B} - 4 \pi {\bf M},
\end{equation}
and the Maxwell equations imply
\begin{equation}
{\bf E} =-\nabla \Phi - {\bf \dot{A}}~,  \quad {\bf B} =\nabla \times {\bf A},
\end{equation}
where the scalar and vector potentials, $\Phi$ and ${\bf A}$, satisfy the equations
\begin{equation}
\Box \Phi = -4\pi \nabla \cdot {\bf P}
\end{equation}
\begin{equation}
\Box {\bf A} = 4\pi ({\bf \dot{P}} + \nabla \times {\bf M}),
\end{equation}
with the Lorentz gauge condition
$
\dot{\Phi} + \nabla \cdot {\bf A}=0.
$

Now, for the  metric (\ref{1}) in particular, it follows that
\begin{equation}
4 \pi {\bf P}=  \mathbb{G}  \cdot {\bf E}
\end{equation}
\begin{equation}
4 \pi {\bf M}= \mathbb{G} \cdot {\bf B},
\end{equation}
where $\mathbb{G}$ is a dyad with components:
\begin{equation}
G_{ab}=\left(
  \begin{array}{ccc}
    f & g & 0 \\
    g & -f & 0 \\
    0 & 0 & 0 \\
  \end{array}
\right)~. \label{G}
\end{equation}

In flat space-time, an electromagnetic plane wave is given by ${\bf E}^{(0)}= \mathbfcal{E} e^{-i\omega t + {\bf
k} \cdot {\bf r}}$ and ${\bf B}^{(0)}= \mathbfcal{B} e^{-i\omega t + {\bf k} \cdot {\bf r}}$, where
$\mathbfcal{E}$ and $\mathbfcal{B}$ are constant vectors such that
\begin{equation}
\omega \mathbfcal{B} = {\bf k} \times \mathbfcal{E}, \quad \omega \mathbfcal{E} = -{\bf k} \times \mathbfcal{B},
\label{EB}
\end{equation}
${\bf k}$ is the wave vector and $ \omega = |{\bf k}| $ the frequency of the wave. The important point is that, if
terms of second order in $G_{ab}$ are neglected, ${\bf P}$ and ${\bf M}$ depend only on the unperturbed electric
and magnetic fields, ${\bf E}^{(0)}$ and ${\bf B}^{(0)}$, and, accordingly, we can set
\begin{equation}
4\pi {\bf P}= \mathbb{G} \cdot \mathbfcal{E} ~e^{-i\omega t + i{\bf k} \cdot {\bf r}}
\end{equation}
\begin{equation}
4\pi {\bf M}= \mathbb{G} \cdot \mathbfcal{B} ~e^{-i\omega t + i{\bf k} \cdot {\bf r}}.
\end{equation}

It is now convenient to define
$$
h_{\pm} (u) = f (u) \mp i g(u),
$$
so that
\begin{equation}
\mathbb{G} \cdot \mathbfcal{E} = h_+ (u) {\cal E}_+ {\bf e}_+  +h_- (u) {\cal E}_- {\bf e}_-
\end{equation}
\begin{equation}
\mathbb{G} \cdot \mathbfcal{B} = h_+ (u) {\cal B}_+ {\bf e}_+  +h_- (u) {\cal B}_- {\bf e}_-~,
\end{equation}
where
$$
{\bf e}_{\pm} = {\bf e}_x \pm i {\bf e}_y
$$
and
$$
{\cal E}_{\pm} = \frac{1}{2} ({\cal E}_x \pm i {\cal E}_y )
$$
$$
{\cal B}_{\pm} = \frac{1}{2} ({\cal B}_x \pm i {\cal B}_y ).
$$

Setting the first order corrections to the potentials in the forms
$$
\Phi^{(1)} = \phi (u) ~e^{-i\omega t + i{\bf k} \cdot {\bf r}},
$$
$$
{\bf A}^{(1)} = {\bf {\cal A}} (u) ~e^{-i\omega t + i{\bf k} \cdot {\bf r}},
$$
it follows that
\begin{equation}
\Box \Phi^{(1)} = -2i (\omega - k_z) \phi ~'(u) ~e^{-i\omega t + {\bf k} \cdot {\bf r}}= -4\pi \nabla \cdot {\bf P}, \label{da1}
\end{equation}
\begin{equation}
\Box {\bf A}^{(1)} = -2i (\omega - k_z) {\bf {\cal A}}~' (u) ~e^{-i\omega t + {\bf k} \cdot {\bf r}}=4\pi ({\bf \dot{P}} + \nabla \times {\bf M}),\label{da2}
\end{equation}
where the primes denote derivation with respect to $u$. These last equations can be integrated separating $+$ and
$-$ components:
$$
\phi^{(1)}= \phi_+^{(1)} + \phi_-^{(1)},
$$
$$
{\bf A}^{(1)}= {\bf A}_+^{(1)} + {\bf A}_-^{(1)}.
$$
It follows that
\begin{equation}
\phi^{(1)}_{\pm} = \frac{1}{(\omega -k_z)} k_{\pm} {\cal E}_{\pm} H_{\pm}  e^{-i\omega t +i {\bf k}\cdot {\bf r}},
\end{equation}
and
$$
{\bf A}^{(1)}_{\pm}= \frac{1}{(\omega -k_z)}\Big\{ \frac{i}{2} \big[{\cal E}_{\pm} (H'_{\pm} - i\omega H_{\pm})
\pm i {\cal B}_{\pm} (H'_{\pm} - i k_z H_{\pm}) \big] {\bf e}_{\pm}
$$
\begin{equation}
\mp i {\cal B}_{\pm}  k_{\pm} H_{\pm}{\bf e}_z \Big\} e^{-i\omega t +i {\bf k}\cdot {\bf r}},
\end{equation}
where we have defined
$$
H'_{\pm}(u)=h_{\pm}(u)
$$
and
$$
k_{\pm} = \frac{1}{2} (k_x \pm i k_y).
$$

Accordingly, the first order correction to the electric field vector can be written as the sum of two terms, ${\bf
E}^{(1)}_+$ and ${\bf E}^{(1)}_-$, such that
\begin{equation} {\bf E}^{(1)}_{\pm}= \big( {\cal E}_{\pm} {\bf
M}_{\pm} \pm i{\cal B}_{\pm} {\bf N}_{\pm} \big) e^{-i\omega t +i {\bf k} \cdot {\bf r}},
\end{equation}
where
$$ {\bf M}_{\pm} \equiv M_{\mp} {\bf e}_{\pm} + M_{\pm z} {\bf e}_z  - \frac{ik_{\pm}}{\omega - k_z}
H_{\pm} {\bf k},
$$
$$
{\bf N}_{\pm} \equiv N_{\mp} {\bf e}_{\pm} + N_{\pm z} {\bf e}_z ,
$$
with
\begin{equation}
M_{\mp} = - ~\frac{i}{2(\omega - k_z)}  \big( H''_{\pm} -2i\omega
H'_{\pm} -\omega^2 H_{\pm} \big)~,
\end{equation}
\begin{equation}
M_{\pm z} =  \frac{k_{\pm}}{\omega - k_z}  H'_{\pm},
\end{equation}
and
\begin{equation}
N_{\mp} = - ~\frac{i}{2(\omega - k_z)}  \big( H''_{\pm} -i(\omega + k_z)
H'_{\pm} -\omega k_z H_{\pm} \big)~,
\end{equation}
\begin{equation}
N_{\pm z} =  \frac{k_{\pm}}{\omega - k_z}  (H'_{\pm} - i\omega H_{\pm} ).
\end{equation}

Define now two orthonormal vectors perpendicular to ${\bf k}$:
$$
 \boldsymbol\epsilon_1 = \frac{1}{k_{\bot}} {\bf e}_z \times {\bf k},
$$
\begin{equation}
 \boldsymbol\epsilon_2 =  \frac{1}{\omega k_{\bot}} (\omega^2 {\bf e}_z  - k_z {\bf k}),\label{ep}
\end{equation}
where $k_{\bot} = (k_x^2 + k_y^2)^{1/2}$, and also a circular polarization basis, which is conveniently chosen as
\begin{equation}
 \boldsymbol\epsilon_{\pm} =  \boldsymbol\epsilon_2 \pm i  \boldsymbol\epsilon_1.\label{en}
\end{equation}

The matrix of the Stokes parameters, as defined in general in the Appendix, can be written in the form $\mathbb{S} + \Delta
\mathbb{S}$, where $\mathbb{S}$ is the corresponding matrix in flat space-time and $\Delta \mathbb{S}$ is the
first order correction produced by the gravitational wave. Explicitly:
\begin{equation}
 \Delta \mathbb{S} =\left(
   \begin{array}{c}
     \boldsymbol\epsilon_+ \cdot {\bf E}^{(1)}  \\
     \boldsymbol\epsilon_- \cdot {\bf E}^{(1)} \\
   \end{array}
 \right)
\left(
  \begin{array}{cc}
    (\boldsymbol\epsilon_+ \cdot {\bf E}^{(0)})^*~, & (\boldsymbol\epsilon_- \cdot {\bf E}^{(0)})^*  \\
  \end{array}
\right) + {\rm h.~ c.}
\end{equation}
Setting  $\Delta \mathbb{S} \equiv \Delta\mathbb{S}_+ +
\Delta\mathbb{S}_-$ and using Eqs. (\ref{SEE}) and (\ref{EE}) in the Appendix, it follows with some straightforward matrix algebra that
\begin{equation}
\Delta \mathbb{S}_{\pm} = \frac{1}{2} \begin{pmatrix}
  \boldsymbol\epsilon_+ \cdot {\bf e}_{\pm} & \boldsymbol\epsilon_+ \cdot {\bf e}_z \\
  \boldsymbol\epsilon_- \cdot {\bf e}_{\pm} & \boldsymbol\epsilon_- \cdot {\bf e}_z
\end{pmatrix}
\begin{pmatrix}
   M_{\mp} & N_{\mp} \\
  M_{\pm z} & N_{\pm z}
\end{pmatrix}
  \begin{pmatrix}
  \boldsymbol\epsilon_- \cdot {\bf e}_{\pm} & \boldsymbol\epsilon_+ \cdot {\bf e}_{\pm} \\
  \pm \boldsymbol\epsilon_- \cdot {\bf e}_{\pm} & \mp \boldsymbol\epsilon_+ \cdot {\bf e}_{\pm}
\end{pmatrix}
\mathbb{S}  + {\rm h. c.},\label{bla3}
\end{equation}
where, according to our previous definitions (\ref{ep}) and (\ref{en}),
\begin{eqnarray}
\boldsymbol\epsilon_+ \cdot {\bf e}_{\pm}  &=& 2\frac{\mp \omega - k_z}{\omega k_{\bot}}~k_{\pm} ,  \nonumber \\
\boldsymbol\epsilon_- \cdot {\bf e}_{\pm}&=& 2\frac{\pm \omega - k_z}{\omega k_{\bot}}~k_{\pm} ,  \nonumber \\
\boldsymbol\epsilon_{\pm} \cdot {\bf e}_z &=& \frac{ k_{\bot}}{\omega }~.
\end{eqnarray}

In particular, we can choose without loss of generality the coordinates system such that the vector ${\bf k}$ lies
in the $(x,z)$ plane, that is $k_y=0$ and $k_{\pm}= \frac{1}{2} k_x$. In this case, Eq. (\ref{bla3}) takes the
simpler form:
\begin{equation}
\Delta \mathbb{S}_{\pm} = \frac{1}{2\omega^2 } \begin{pmatrix}
  \mp \omega -k_z & k_x \\
  \pm \omega -k_z & k_x
\end{pmatrix}
\begin{pmatrix}
   M_{\mp} & N_{\mp} \\
  M_{\pm z} & N_{\pm z}
\end{pmatrix}
  \begin{pmatrix}
  \pm \omega - k_z & \mp \omega - k_z \\
  \omega \mp k_z & \omega \pm k_z
\end{pmatrix}
\mathbb{S}  + {\rm h. c.}\label{bla}
\end{equation}

\section{Constant amplitude gravitational wave}

As an example of application of the general formula given above, consider a constant amplitude sinusoidal
gravitational wave, such as one generated by a periodically varying configuration of massive bodies (see, e.g.,
Landau and Lifshitz \cite{LL}). Accordingly we set
\begin{equation}
H_{\pm} =h_0 e^{\mp i \Omega u \mp i \alpha} ,
\end{equation}
where $h_0$ is a real valued constant, $\Omega$ is the frequency of the wave and $\alpha$ is a constant phase. In
this particular case:
\begin{equation}
M_{\mp} = \frac{i}{2(\omega - k_z)}  (\Omega \pm \omega)^2 H_{\pm} ~,
\end{equation}
\begin{equation}
M_{\pm z} = \mp i k_{\pm}~ \frac{\Omega}{\omega - k_z}  H_{\pm},
\end{equation}
and
\begin{equation}
N_{\mp} = \frac{i}{2(\omega - k_z)}  (\Omega \pm \omega) (\omega \pm  k_z) H_{\pm} ~,
\end{equation}
\begin{equation}
N_{\pm z} = - i k_{\pm}~ \frac{\omega \pm \Omega}{\omega - k_z}  H_{\pm}~.
\end{equation}

Now, in most practical cases $\Omega \ll \omega$ and, accordingly, terms of order $\Omega/\omega$ can be
neglected. In this case, the above equations further simplify to
\begin{equation}
M_{\mp} = \frac{i \omega^2}{2(\omega - k_z)}   H_{\pm} ~,
\end{equation}
\begin{equation}
M_{\pm z} = 0,
\end{equation}
and
\begin{equation}
N_{\mp} = \pm \frac{i}{2(\omega - k_z)}  \omega (\omega \pm  k_z) H_{\pm} ~,
\end{equation}
\begin{equation}
N_{\pm z} = - i k_{\pm}~ \frac{\omega }{\omega - k_z}  H_{\pm}~.
\end{equation}
After substitution in Eq. (\ref{bla}), the first order correction to the Stokes parameters turns out to be
$$
\Delta \mathbb{S}=\Delta \mathbb{S}_+ +\Delta \mathbb{S}_- = - \frac{i}{4(\omega - k_z)}
$$
\begin{equation}
\times
\begin{pmatrix}
  3 k_x^2 (H_+ + H_-) & (\omega + k_z)^2 H_+ +(\omega - k_z)^2 H_- \\
  -(\omega - k_z)^2 H_+ -(\omega + k_z)^2 H_- & k_x^2 (H_+ + H_-)
\end{pmatrix}
\mathbb{S} + {\rm h.c.}\label{bla2}
\end{equation}

Now, in the particularly important case of unpolarized light, the averaged Stokes parameters are
$$
\big< s_i \big>=0, ~~i=1,2,3
$$
and $\big< s_0 \big>$ is just the intensity of the wave. In this case, it follows from Eq. (\ref{bla2}) that
\begin{equation}
\Delta \big< s_0 \big> =0, \quad  \Delta \big< s_3 \big> =0,
\end{equation}
and
\begin{equation}
\big< s_1 \big> + i  \big< s_2 \big> = \big< s_0 \big> \frac{h_0}{2(\omega - k_z)} \Big[2\omega k_z \sin \theta +
i (\omega^2 + k_z^2) \cos \theta \Big],\label{s12}
\end{equation}
where $\theta = \Omega u + \alpha$. These are precisely the conditions for a light beam to be linearly polarized
(as can be seen, for instance, from the definition of the Poincar\'e sphere; see, e.g., Born and Wolf \cite{BW}).

\section{Concluding remark}

The main result of this article is the formula given by Eq. (\ref{bla3}), or its simplified form (\ref{bla}). This
formula is based on the general expressions for the electromagnetic potentials and electric field given in Section
2, and it permits to calculate the Stokes parameters of light in the presence of a gravitational wave. An
application of the present formalism to the case of a constant amplitude gravitational wave shows that an
unpolarized electromagnetic wave acquires a linear polarization, with the direction of polarization varying in
synchrony with the gravitational wave. This result is consistent with the one obtained in Ref. \cite{hac}.

\section*{Acknowledgments}
Work supported in part by PAPIIT-UNAM, project IN 101511-3.

\section*{Appendix A: Stokes parameters}

\renewcommand{\theequation}{\Alph{section}.\arabic{equation}}

\setcounter{section}{1} \setcounter{equation}{0}

Given the electric field ${\bf E}$ of a plane electromagnetic wave propagating in the ${\bf k}$ direction, the polarization can be described by the
Stokes parameters constructed from the two scalar products ${\bf E} \cdot \boldsymbol\epsilon_{\pm}$, where $
\boldsymbol\epsilon_{\pm}$ are two complex vectors defined by (\ref{ep}) and (\ref{en}). Superscript $(0)$ for
the unperturbed field are dropped in the present appendix.

The Stokes parameters are defined as
\begin{eqnarray}
s_0 &=& \frac{1}{2} (| \boldsymbol\epsilon_+ \cdot {\bf E} |^2  + | \boldsymbol\epsilon_- \cdot {\bf E} |^2 ) \nonumber \\
s_1 +i s_2 &=&   ( \boldsymbol\epsilon_+ \cdot {\bf E})^* ( \boldsymbol\epsilon_- \cdot {\bf E}) \nonumber \\
s_3 &=&\frac{1}{2} ( | \boldsymbol\epsilon_+ \cdot {\bf E} |^2  - | \boldsymbol\epsilon_- \cdot {\bf E} |^2)~,
\end{eqnarray}
following the notation of Jackson\cite{jac} (except for a factor $ \sqrt{2}$ in the definition of
$\boldsymbol\epsilon_{\pm}$). This can be written in matrix form as
\begin{equation}
\mathbb{S} \equiv \left(
  \begin{array}{cc}
    s_0+s_3 & s_1-is_2 \\
    s_1+is_2 & s_0 -s_3 \\
  \end{array}
\right) = \left(
   \begin{array}{c}
     \boldsymbol\epsilon_+ \cdot {\bf E}  \\
     \boldsymbol\epsilon_- \cdot {\bf E} \\
   \end{array}
 \right)
\left(
  \begin{array}{cc}
    (\boldsymbol\epsilon_+ \cdot {\bf E})^*~, & (\boldsymbol\epsilon_- \cdot {\bf E})^*  \\
  \end{array}
\right).
\end{equation}

Using the relations $\omega {\bf B}= {\bf k} \times {\bf E}$ and $\omega {\bf E}= -{\bf k} \times {\bf B}$ in
combination with (\ref{ep}) and (\ref{en}), it follows that
\begin{equation}
\boldsymbol\epsilon_{\pm} \cdot {\bf E} = \frac{\omega}{k_{\bot}} (E_z \pm i B_z),
\end{equation}
 and therefore
\begin{equation}
\mathbb{S}   = \frac{2\omega^2}{k_{\bot}^2} \left(
   \begin{array}{c}
     (E_z + i B_z)  \\
     (E_z - i B_z) \\
   \end{array}
 \right)
\left(
  \begin{array}{cc}
    (E_z + i B_z)^*~, & (E_z - i B_z)^*  \\
  \end{array}
\right). \label{SEE}
\end{equation}

Also
\begin{eqnarray}
{\bf E} &=& \frac{\omega}{2k_{\bot}} [(E_z -i B_z) \boldsymbol\epsilon_+ + (E_z +i B_z) \boldsymbol\epsilon_-], \nonumber \\
i{\bf B} &=& \frac{\omega}{2k_{\bot}} [-(E_z -i B_z) \boldsymbol\epsilon_+ + (E_z +i B_z) \boldsymbol\epsilon_-] ,
\end{eqnarray}
and since
$${\bf E}= E_- {\bf e}_+ + E_+ {\bf e}_- + E_z {\bf e}_z,$$
with a similar expression for ${\bf B}$, it follows that
\begin{equation}
\begin{pmatrix}
  E_{\pm} \\
  \pm i B_{\pm}
\end{pmatrix}
= \frac{\omega}{4k_{\bot}} \begin{pmatrix}
  \boldsymbol\epsilon_- \cdot {\bf  e}_{\pm} & \boldsymbol\epsilon_+ \cdot {\bf  e}_{\pm} \\
  \pm \boldsymbol\epsilon_- \cdot {\bf  e}_{\pm} & \mp \boldsymbol\epsilon_+ \cdot {\bf  e}_{\pm}
\end{pmatrix}
\begin{pmatrix}
  E_z + i B_z \\
  E_z - i B_z
\end{pmatrix}.\label{EE}
\end{equation}


\begin{thebibliography}{99}

\bibitem{ligo} LIGO project: http://www.ligo.caltech.edu/

\bibitem{rus}  A. A. Starobinskii, Sov. Astr. Lett. {\bf 11}, 133 (1985). A. G. Polnarev, Sov. Astr. {\bf 29}, 607
(1985).

\bibitem{cosm} A. C. S. Readhead {\it et al.}, Science {\bf 306}, 836 (2004).

%DOI: 10.1126/science.1105598.

\bibitem{pleb} J. Plebanski, Phys. Rev. {\bf 118}, 1396 (1960).

\bibitem{sib} N. R. Sibgatullin, {\it Oscillations and Waves: In Strong Gravitational and Electromagnetic Fields},
Springer-Verlag (Berlin, 1991). Section 1.3

\bibitem{ehler} J. Ehlers and W. Kundt, in {\it The Theory of Gravitation}, L. Witten, editor, John Wiley \& Sons, Inc., (New York and
London, 1962); p. 86-101.

\bibitem{hac} S. Hacyan, Gen. Rel. Grav. {\bf 44}, 2923 (2012).
%DOI 10.1007/s10714-012-1434-4

\bibitem{LL} L. D. Landau and E. M. Lifshitz, {\it The Classical Theory of Fields} (4th edition), Butterworth-Heinemann,
(Oxford, 2000). Chap. 13.
%\bibitem{mtw} C. W. Misner, K. S. Thorne, and J. A. Wheeler, {\it Gravitation}, W. H. Freeman, San Francisco 1973; Sect. 22.5.

\bibitem{jac} J. D. Jackson, {\it Classical Electrodynamics}. Wiley; 2 edition (Neew York, London, 1975). Sect. 7.2.

\bibitem{BW} M. Born and E. Wolf, {\it Principles of Optics}. Pergamon Press, (Oxford, 1975); Sect. 30-31.





\end{thebibliography}
\end{document}